\documentclass[11pt, onecolumn]{article} 
\usepackage{graphicx,import}
\usepackage{caption}
\usepackage[final]{changes}
\definechangesauthor[name=(R1.1), color=red]{(Rev1.1)}
\definechangesauthor[name=(R1.2), color=red]{(Rev1.2)}
\definechangesauthor[name=(R1.3a), color=red]{(Rev1.3a)}
\definechangesauthor[name=(R1.3b), color=red]{(Rev1.3b)}
\definechangesauthor[name=(R1.3c), color=red]{(Rev1.3c)}
\definechangesauthor[name=(R1.4), color=red]{(Rev1.4)}
\definechangesauthor[name=(R2.2), color=red]{(Rev2.2)}
\definechangesauthor[name=(R2.3), color=red]{(Rev2.3)}
\definechangesauthor[name=(R2.4), color=red]{(Rev2.4)}
\definechangesauthor[name=(R2.5), color=red]{(Rev2.5)}
\definechangesauthor[name=(R2.6), color=red]{(Rev2.6)}
\definechangesauthor[name=(R2.7), color=red]{(Rev2.7)}
\definechangesauthor[name=(R2.9), color=red]{(Rev2.9)}
\definechangesauthor[name=(R2.10), color=red]{(Rev2.10)}
\definechangesauthor[name=(R2.11), color=red]{(Rev2.11)}
\definechangesauthor[name=(R2.12), color=red]{(Rev2.12)}
\definechangesauthor[name=(R2.13), color=red]{(Rev2.13)}
\definechangesauthor[name=(R2.14), color=red]{(Rev2.14)}
\definechangesauthor[name=(R1-R1), color=blue]{(Rev1-R1)}
\definechangesauthor[name=(R2-R1), color=blue]{(Rev2-R1)}

\let\oldmarginpar\marginpar
\renewcommand\marginpar[1]{\-\oldmarginpar[\raggedleft\scriptsize #1]%
{\raggedright\footnotesize #1}}
\begin{document}

\title{Perfusion and apparent oxygenation in the human placenta (PERFOX)}

\author{\begin{minipage}{5in} \begin{center}    
Jana Hutter$^{\star \diamond}$, Anita A. Harteveld $^{+}$,Laurence H. Jackson$^{\star \diamond}$,Suzanne Franklin$^{+}$, Clemens Bos$^{+}$, Matthias J. P. van Osch$^{*}$, Jonathan O'Muircheartaigh$^{\star \diamond}$,Alison Ho$^{\circ}$, Lucy Chappell$^{\circ}$, Joseph V Hajnal$^{\star \diamond}$,Mary Rutherford$^{\star \diamond}$,Enrico De Vita$^{\diamond}$
\end{center} 
\end{minipage}}
\maketitle
\noindent Affiliations:\\
$^{\star}$ Centre for the Developing Brain, King's College London, United Kingdom \\
$^{\diamond}$ Biomedical Engineering Department, King's College London, United Kingdom \\
$^{+}$ Department of Radiology, University Medical Center Utrecht. Utrecht, The Netherlands\\
$^{*}$C.J. Gorter Center for high field MRI, Department of Radiology, Leiden University Medical Center, Leiden, The Netherlands\\
$^{\circ}$ Women's Health Academic Centre, King's College London, London, United Kingdom
\\
\thispagestyle{empty}
\small
Running Head: PERFOX\\
\\
Correspondence to: Jana Hutter, Centre for the Developing Brain, King's College London, 1st Floor South Wing, St. Thomas Hospital, SE1 7EH, London, United Kingdom. E-mail: jana.hutter@kcl.ac.uk\\
Total word count: 4250, 9 figures, 0 tables, 5 Supporting Information Figures\\
\\


\begin{abstract}
\noindent \textbf{Purpose:} To study placental function - both perfusion and an oxygenation surrogate ( T$_2^*$) -  simultaneously and quantitatively in-vivo.\\
\textbf{Methods:} 15 pregnant women were scanned on a 3T MR scanner. For perfusion measurements, a velocity selective arterial spin labelling preparation module was placed before a multi-echo gradient echo EPI readout to integrate T2* and perfusion measurements in one joint perfusion-oxygenation (PERFOX) acquisition. Joint motion correction and quantification were performed to evaluate changes in T2* and perfusion over GA.\\
\noindent \textbf{Results:} The optimised integrated PERFOX protocol and post-processing allowed successful visualization and quantification of perfusion and  T$_2^*$ \deleted[id=(Rev1-R1)]{both sequential scans and} in all subjects. Areas of high  T$_2^*$ and high perfusion \deleted[id=(Rev1.1)]{- called in the following  T$_2^*$ or respectively perfusion peaks -} appear to correspond to placental sub-units and show a systematic offset in location along the maternal-fetal axis. \added[id=(Rev1.1)]{The areas of highest perfusion} \added[id=(Rev1-R1)]{are consistently closer to the maternal basal plate and the} \added[id=(Rev1.1)]{areas of highest  T$_2^*$} closer to the fetal chorionic plate.\deleted[id=(Rev1-R1)]{A consistent shift of high-perfusion areas vs high- T$_2^*$ areas is apparent, with high perfusion consistently more towards the basal plate and high  T$_2^*$ towards the chorionic plate.} Quantitative results show a strong negative correlation of gestational age with T$_2^*$ and weak negative correlation with perfusion.\\

\noindent \textbf{Conclusion:} \deleted[id=(Rev2.14)]{Concomitant measurement of placental  T$_2^*$ and perfusion was successful; the compromise of having a reduced number of echoes and slices is compensated by the joint sequence's ability to provide truly simultaneous and co-registered estimates of local  T$_2^*$ and perfusion.} \added[id=(Rev2.14)]{A strength of the joint sequence is that it provides truly simultaneous and co-registered estimates of local  T$_2^*$ and perfusion, however, to achieve this, the time per slice is prolonged compared to a perfusion only scan which can potentially limit coverage.} The achieved interlocking can be particularly useful when quantifying transient physiological effects such as uterine contractions. PERFOX opens a new avenue to elucidate the relationship between maternal supply and oxygen uptake, both of which are central to placental function and dysfunction.\\


\end{abstract}

\section*{Introduction}
\begin{figure}[t]
  \centering
  \scriptsize
  \includegraphics[width=0.6\textwidth]{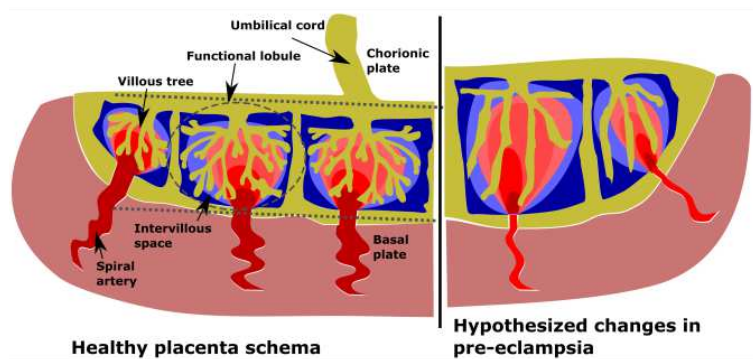}
  \caption{ The placenta is depicted schematically. The spiral arteries ensuring supply from maternal side, the villous trees for oxygen uptake as well as a (schematic) depiction of the oxygen concentration from red (high oxygen content) to blue (low oxygen content) are illustrated together with the definition for a functional lobule used in the following. Furthermore, dotted lines point to the basal and chorionic plate. The right side shows some of the hypothesized changes in pre-eclampsia: elongated, less-capillarized trees and increased thickness.}
  \label{fig:placenta}
\end{figure}

The human placenta constitutes the only link between mother and fetus. It supplies the fetus with oxygen and nutrients and ensures the elimination of waste products. It furthermore has essential endocrine and immunological functions. Major pregnancy complications such as pre-eclampsia, fetal growth restriction and preterm birth - all carrying a substantial risk of increased morbidity and mortality for both mother and child - are linked with placental insufficiency. Anatomically, the human placenta is composed of 10-40 functional lobules \cite{Burton2009}. Each lobule contains 1-2 maternal spiral arteries, suppling maternal blood from the uterine arteries into the inter-villous space and thus irrigating these exchange units with oxygen-rich blood (see Fig. \ref{fig:placenta}, left side). The fetal villi contain extensive arterial-capillary-venous systems originating from the umbilical cord, and are bathed in maternal blood. Special adaptions in early pregnancy include remodelling of the maternal spiral arteries, which allows for a slow and constant blood flow and thus ideal perfusion of the functional units by maximizing the contact area and the transfer time. Ongoing maturation across gestation leads to denser, more capillarized vascular trees and decreasing thickness of the cellular layer which separates villi and maternal blood.\\

\noindent Placental insufficiency is linked to smaller then normal vascular lumen of the maternal spiral arteries \added[id=(Rev2.14)]{as a sign of incomplete remodelling}. Furthermore, less capillarized, elongated villi can be observed \cite{Burton2009} and are schematically depicted in Fig. \ref{fig:placenta} (right side). These structural findings are generally characterized ex-vivo using histopathology. While these insights are valuable, they do not directly inform on causality or the cascade of events that might link different structural features. This requires imaging the placenta in-vivo, which is currently mostly performed using ultrasound (US). \deleted[id=(Rev2.14)]{while it is functioning.} However, current US screening tends to focus mainly on flow measurements in the umbilical cord and uterine artery and thus fails to provide direct insight into the placental functional core. \deleted[id=(Rev2.14)]{and cascade of events within the placenta.} This lack of a suitable in-vivo observation window hampers early diagnosis and prevents understanding of the complex disease aetiology.\\

\noindent A recent surge in placental MRI studies showed promising results to bridge this gap and allow in-vivo assessment of placental function during gestation. The ability of MRI to generate contrasts adapted to microstructure and tissue properties renders this technique ideally suited to visualize the cascade of events within the placenta leading ultimately to placental insufficiency. Among these, perfusion measurements have been performed using a variety of techniques:  Intra-voxel-incoherent motion (IVIM) \cite{Bonel2010,Moore2000,Slator2018,Hutter2018,Sohlberg2014,Moore2000,Manganaro2010,Jakab2017,Melbourne2018,Siauve2017}, Arterial Spin Labelling (ASL) \cite{Duncan1998,Francis1998,Gowland1998,Derwig2013,Ludwig2018}, and lately, Velocity-selective Arterial Spin Labeling (VSASL) \cite{Zun2017,Zun2018} have all been used. VSASL has the advantage that it does not require a geometrical separation of the blood labelling region and the perfusion observation region. It labels blood that is flowing at a speed above a user defined cutoff. The longitudinal magnetization of blood flowing above this cutoff is saturated. During a post-label delay the tagged blood flows down the arterial system and modifies the magnetization in the imaging volume.\\

\noindent Relaxometry techniques has been successfully used for oxygenation studies of the placenta \cite{Sinding2017,Sorensen2013,Derwig2013a,Ingram2016,Huen2013,Hutter2018}. \deleted[id=(Rev2.14)]{T$_2^*$ relaxometry has been successfully used for oxygenation studies of the placenta.}\added[id=(Rev2.2)]{ T$_2^*$ is of specific interest because there is a well established relationship between this parameter and oxygenation through the Blood Oxygen Level Dependent (BOLD) \cite{Ogawa1990}. Whilst it is reasonable to regard T$_2^*$ as an indicator of oxygen concentration it is not a direct measure. Important confounding factors include micro-structural geometry effects, e.g. due to the random diffusion of water molecules around vessels which lead to a reduced BOLD signal around smaller vessels and differences in the oxygen-hemoglobin dissociation curve between fetal and adult hemoglobin.} \deleted[id=(Rev2.2)]{It constitutes a flow-independent measurement of tissue properties on the voxel scale \cite{Sorensen2013}. The changes in magnetic susceptibility when diamagnetic oxyhemoglobin releases its oxygen are known as the BOLD effect \cite{Ogawa1990}. The resulting deoxyhaemoglobin is paramagnetic, and the produced phase dispersion reduces the signal intensity at the time of the echo, leading to a decrease in T2$^*$ values. Thus, T$_2^*$ is related to the total amount of deoxyhaemoglobin in the voxel and high T$_2^*$ values indicate high oxygen concentrations. It is, however, not a linear relationship, as other sources of field inhomogeneities originating from microscopic interactions between neighboring atoms give rise to additional signal loss.}

\added[id=(Rev2.3)]{Despite the recent increase in available techniques and interest, current placental imaging studies are often limited by focusing on an individual contrast and a tendency to evaluate parameters averaged over the entire placental volume. Given the complex disease aetiology, physiological placenta studies could benefit from multi-parametric analyses, that can locally link e.g. maternal perfusion to the microscopic structure of the villous tree and oxygen exchange.} There are, however, two significant challenges complicating such described multi-modal assessments in placental MRI : (i) \deleted[id=(Rev2.14)]{motion from } maternal respiration and fetal bulk movements decrease internal consistency between data acquistions separated in time \deleted[id=(Rev2.14)]{and call for post-processing correction methods;} (ii) examination times need to be kept short to ensure maternal comfort \deleted[id=(Rev2.14)]{during the MRI scan limits the possible acquisition time.}\added[id=(Rev2-R1)]{This study therefore proposes a multi-dimensional simultaneous integrated assessment of perfusion and oxygenation called PERFOX: combining two independent functional MRI techniques,  T$_2^*$ relaxometry and VSASL, allows to study the interaction between \added[id=(Rev2.14)]{maternal perfusion and T$_2^*$ as a marker fetal oxygen uptake}. Quantitative and qualitative results from 15 placentas illustrate the dynamic joint spatial patterns of perfusion and oxygenation in-vivo \deleted[id=(Rev2.14)]{as well as trends }over gestational age.}

\section*{Methods}

\subsection*{\added[id=(Rev2.14)]{Experiments}}
\noindent The study was approved by the local IRB (Riverside Ethics Committee REC 14/LO/1457). A total of 15 pregnant women (median/range gestational age (GA) 28.9/21.9-38.2 weeks) were included and scanned subsequent to informed consent, in the supine position \cite{Hughes2016}  \added[id=(Rev2.14)]{on a clinical 3T Philips Achieva MRI scanner (Best, Netherlands) with a 32-channel receiver coil.} \added[id=(Rev2-R1)]{Safety and comfort of the mother was ensured: bespoke padding was designed to support the lower back, all women were asked to lie on the left side first to shift the weight of the pregnant uterus off the vena cava before slowly transitioning into supine position. Furthermore, life monitoring using constant pulse oximetry and blood pressure measurements were performed at 10-minute intervals, and the scanner operator maintained frequent verbal interaction with the women. The examination was split in two sessions of 30min separated by a break to increase patient comfort.} \added[id=(Rev1.2)]{All the women scanned for this study tolerated the supine position well.}
\subsection*{\deleted[id=(Rev2.14)]{Experiments}}
\deleted[id=(Rev2.14)]{All acquisitions were performed on a clinical 3T Philips Achieva MRI scanner with a 32-channel cardiac receiver coil. }
Each scan session started with initial calibration scans: a T2-weighted 2D single shot Turbo Spin Echo sequence and B0 map were acquired in coronal orientation covering the whole uterus. These enabled both image-based shimming \cite{Gaspar2018} targeted to the placental parenchyma and planning of the acquisition geometry for the functional acquisitions. The proposed PERFOX scan was acquired next.

\noindent \deleted[id=(Rev2.14)]{In the following, general details of the PERFOX acquisition read-out, contrast mechanisms and post-processing are illustrated and the acquired data and experiments described in detail.}

\subsection*{PERFOX Read-out}
\noindent Strategies to deal with motion is of key importance due to the high prevalence of breathing and fetal motion. The image acquisition was thus performed with single-shot echo planar imaging (ssEPI) to freeze motion within each slice. \deleted[id=(Rev2.14)]{and thus to obtain high-quality data.} To limit acoustic noise, the ssEPI sequence was constrained by imposing an echo spacing of ~1ms (i.e. read-out frequency of ~500Hz), shown previously to minimize acoustic noise on our scanner \cite{Hutter2018quepi}. Acoustic noise measurements were performed using an MR-compatible Optoacoustics Fiber Optic Microphone (Optimic 1155, resolution of 0.1 dB) to verify that the acoustic noice was kept below 105dB(A). \added[id=(Rev2.14)]{The sensor was positioned at isocenter, the typical location of the fetal head, in the empty scanner bore to ensure stable measurement conditions.} \deleted[id=(Rev2.14)]{Noise measurements were made with the sensor positioned at isocenter in the empty scanner bore, the typical location of the fetal head. Empty bore measurements were used in order to create stable conditions to allow for comparison across different measurement sessions. These measurements were aimed to verify that the acoustic noise was kept below 105dB (A-weighting).} For most scans, the coronal slice orientation relative to maternal habitus was selected for maximal efficiency as it ensures that the longest dimension placentas located, as is most common, mainly anterior or posterior is parallel to the slice plane. An axial slice orientation was chosen for select acquisitions to better visualise the placenta \added[id=(Rev2.14)]{from maternal to fetal side in one plane}. Both in-plane resolution and slice thickness were fixed to 4mm.\deleted[id=(Rev2.14)]{, however, for a few test scans higher resolution was chosen.the resolution was higher to demonstrate the findings in higher resolution.}

\subsection*{Intrinsic contrast mechanisms}
\noindent VSASL \cite{Wong2006}, implemented in a similar manner to \cite{Schmid2015} was employed to visualize perfusion within the placental parenchyma. The sequence consists of a velocity-selective tagging module, parametrized by the cutoff ($V_c$) and a post-label delay (PLD), a background suppression (BGS) module consisting of two inversion pulses, and the EPI read-out module. \added[id=(Rev2.6)]{As in conventional implementations of VSASL, the tagging module is spatially non-selective. However, the gradients in the tagging module are applied along a chosen axis and only blood flowing in this direction is labelled. Labelling in the maternal superior-inferior direction was judged most effective (largest blood signal change) in preliminary investigations irrespective of scan plane orientation, so this was adopted for all examinations.} Acquisition of control images with the gradients in the tagging module set to zero and subtracting these from labelled images \added[id=(Rev2.14)]{removes the static tissue signal contributions.} \deleted[id=(Rev2.14)]{so that only signal from spins flowing above the cut-off velocity $V_c$ at the time of tagging is visualized when taking the difference between control and label images.} \added[id=(Rev2.14)]{Each pair of label and control images, acquired in interleaved order is referred to as one dynamic.} \deleted[id=(Rev2.14)]{Label and control volumes were acquired in interleaved order. Each label-control pair is referred to as one dynamic.} \deleted[id=(Rev2.14)]{The first dynamic is acquired without this BGS module to provide a pseudo-M0. Three preliminary scans and T1 measurements made using ZEBRA \cite{Hutter2018a} were employed to adjust the timing of the BGS pulses during the post-labelling delay to best reduce static placental signal.} A consequence of using a 2D multi-slice acquisition in combination with VSASL \deleted[id=(Rev1-R1)]{for posterior placentas} \added[id=(Rev2.14)]{is that each slice within the imaging volume is excited at a different time relative to the tagging module.} \added[id=(Rev2.4)]{Each slice thus has its own PLD and presents with } different degrees of BGS.

\begin{figure}[ht]
  \centering
  \scriptsize
  \includegraphics[width=0.8\textwidth]{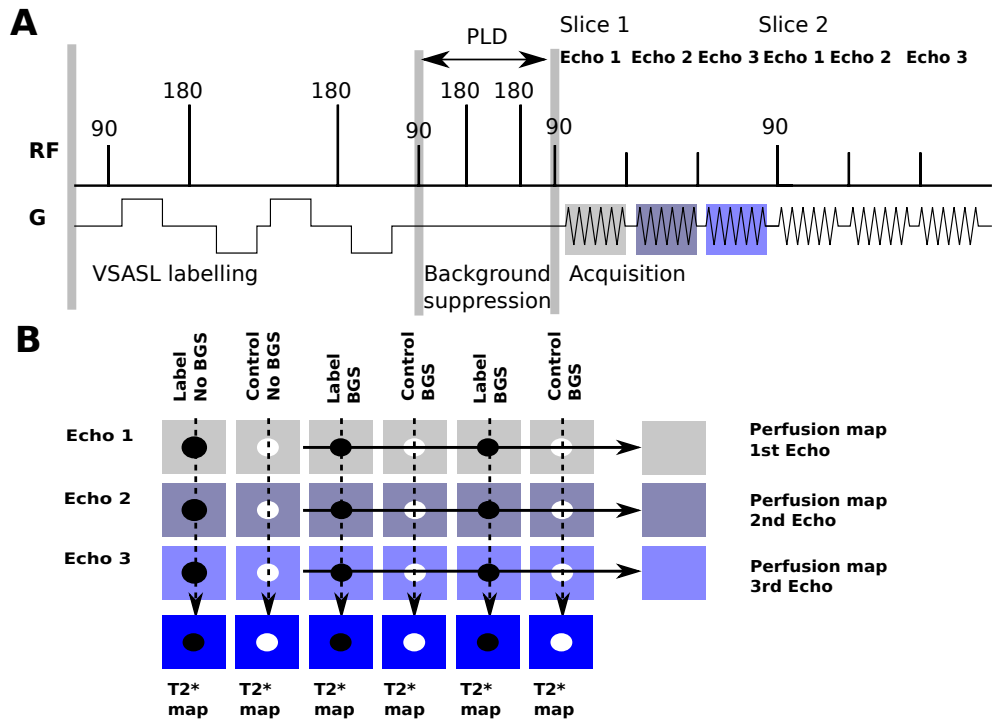}
  \caption{\added[id=(Rev1-R1)]{The acquisition strategy (A) and post-processing (B) for PERFOX are depicted schematically. In (A) the VSASL labelling module, the background suppression module and the acquisition module - composed of three echoes - are depicted. NB. the horizontal time axis is not to scale.}}
  \label{fig:muechos}
\end{figure}

\noindent \added[id=(Rev2.14)]{This basic VSASL sequence was modified and optimized to deliver information on perfusion and  T$_2^*$ simultaneously} by the addition of extra gradient echoes prescribed after the initial echo for each slice. This allows  T$_2^*$ mapping independently for both label and control volumes \deleted[id=(Rev2.14)]{and for all slices} as depicted in Fig. \ref{fig:muechos}a. This approach ensures a reduced sensitivity to motion, as all data required for  T$_2^*$ fitting in each slice is acquired within \textless 200 ms. The range of TEs for the multiple echoes was chosen based on placental  T$_2^*$ values, obtained in  a previous study (10-150ms) \cite{Hutter2018}.\\

\noindent \deleted[id=(Rev2.4)]{perfusion signal due to T1 decay of the labeled blood and T1 recovery of static tissue magnetization.} \deleted[id=(Rev2.4)]{This is important when considering that the main direction of flow for blood within the placenta, following the physiological pathway, is from maternal basal plate to fetal chorionic plate. Therefore for coronal scans, to achieve a consistent PLD variation in this functional direction, the slice acquisition order was adjusted to be in the maternal to fetal direction, such that, according to placental location, the first slice (lowest PLD) was always on the basal plate, i.e.: anterior-to-posterior slice order for anterior placentas and posterior-to-anterior slice order for posterior placentas. In contrast, when imaging in the mother’s transverse orientation, variations of the parametric maps in the functional direction can be captured within each slice, i.e. at the same PLD.}

\added[id=(Rev2.4)]{To assure that the perfusion results are comparable with respect to their position along the axis from maternal basal plate to fetal chorionic plate, the slice acquisition order was adjusted to be anterior-to-posterior for anterior placentas and posterior-to-anterior for posterior placentas.}

Importantly, the inter-slice PLD increment depends on the number of gradient echoes acquired, the higher the number of acquired echoes, the larger the PLD increment. \added[id=(Rev2.11)]{The optimization of the joint scan required adjustment of the number of slices and the number of echoes, whilst ensuring that T$_2^*$ fitting was reliable and the coverage appropriate. Enforcing a similar maximal PLD for the last slice for PERFOX compared to a 'separate' VSASL acquisition resulted in 8 slices, compared to 13 slices (See Supporting Information Figure S1). The intrinsic link between inter-echo spacing and length of the EPI train results in TEs of [20,56,93] ms for the standard PERFOX acquisition.}

\noindent \added[id=(Rev2.11)]{For the BGS, three preliminary T1 experiments made using ZEBRA \cite{Hutter2018a} allowed to determine a T1 range ([900 -1200ms]) which was used to adjust sequence timings to make sure that all imaged slices had positive signal at the beginning of the read-out of the first slice \cite{Vidorreta2014}; this resulted in the two inversion pulses being placed right after the tagging module and 1130ms later. Supporting Information Figure S2 illustrates the effect of the choice of slice orientation on the acquired images, in terms of their differing dependence on PLD.} \added[id=(Rev2.14)]{The first dynamic is acquired without this background suppression module to provide a pseudo-M0.}\\

\added[id=(Rev2.11)]{\noindent The imaging parameters for the coronal PERFOX scans were resolution $4^3mm^3$, Field-of-view 300x380x20mm, 8 slices in ascending order, SENSE 2.5, Partial Fourier 0.97, TR=3500ms, PLD=1600ms, inter-slice spacing 115ms, 1 dynamic without BGS and 25 dynamics with BGS (pulse timings 50ms and 1130ms), Label 50ms, G=13mT/m, V$_c$=2cm/s, total acquisition time 3min.}

\subsection*{Post-processing}
\noindent Nonrigid motion correction was performed in ANTS \cite{AVANTS2008}. \deleted[id=(Rev2.4)]{to align the acquired volumes.} All VSASL volumes of the first echo time were registered to a common representative space created using an iterative template construction approach. After registering the volumes to an initial average of the VSASL dynamics, the template construction algorithm nonlinearly registered each volume to the template image and then constructed a representative shape image requiring the least transformation from all other volumes. This process was repeated but with the new representative image taking the place of the initial average in the registration. The parameters used for the nonlinear registration were the defaults for the script, using the Symmetric Normalization model, and a cost function with a voxel radius of 4. In a second step, the obtained transformations were employed to correct all volumes from subsequent echo times. \added[id=(Rev2.14)]{This two-step registration approach follows the assumption that volumes acquired at the different subsequent TEs are aligned due to their temporal closeness (\textless 200ms) and do not require further registration.} For quantitative whole-organ results, the region of-interest (ROI) was manually drawn on each slice of the first, non-background-suppressed control volume (i.e. the pseudo-M0) .\\

\noindent Once all volumes were aligned, perfusion analysis and  T$_2^*$ fitting were performed as depicted schematically in Fig. \ref{fig:muechos}b. For  T$_2^*$ mapping, the signal values from all echoes were fitted voxel wise to a mono-exponential decay curve $S(t)=S(0)exp^{(-T_2^{*}/TE)}.$ Using Levenberg-Marquart optimization with initial parameters  T$_2^*$=100 [ms] and $S(0)=S(TE_1)$.  Both individual  T$_2^*$ maps for each volume \deleted[id=(Rev2.14)]{, referred to as T$_2*_{d}^{c/t}$} for control/tag volumes respectively, and averaged  T$_2^*$ maps for control and label volumes were calculated.\\

\noindent Perfusion analysis was performed by pairwise subtraction of the label-control pairs with BGS.\deleted[id=(Rev2.14)]{ leading to perfusion-weighted images $P_d^e$ for dynamic $d$ and echo $e$.} \added[id=(Rev1-R1)]{The resulting difference images were normalized by the pseudo-M0. }\deleted[id=(Rev2.4)]{-obtained from the non-background-suppressed control volume.} \added[id=(Rev1-R1)]{This normalisation takes the unsaturated magnetisation of the inflowing blood into account and removes the effect of the T2 weighting of the labelling module on the perfusion data; it therefore produces a semi-quantitative quantity related to blood-flow facilitating comparison of data from different subjects. This avoids absolute quantification which would require estimation or assumption of blood T1 and T2, since these are highly dependent on blood oxygenation, hematocrit and whether maternal or fetal blood is being considered.} We therefore express the \deleted[id=(Rev2.9)]{'quantitative'} perfusion maps \deleted[id=(Rev2.14)]{$\tilde{P_d^e}$}in arbitrary units. All displayed results - both the maps and the quantitative results are using these units. By averaging over multiple dynamics the cumulative result is obtained. \deleted[id=(Rev2.14)]{into account}
\deleted[id=(Rev2.10)]{$P_D^e=1/D \sum_{d=1}^{d=D}\tilde{P_d^e}.$}

\noindent \added[id=(Rev1.4)]{Quantitative evaluation was performed by averaging the values within the ROI to assess the relationship between  T$_2^*$ and perfusion/M0 over GA. A one-way ANCOVA was conducted to determine whether there is a statistically significant difference between placental location (anterior/posterior) on  T$_2^*$ respectively on perfusion/M0 when controlling for gestational age.}\\

\noindent All described analysis so far included the cumulative averaged subtraction results from all dynamics. \added[id=(Rev1-R1)]{The influence of the number of dynamics on the cumulative perfusion map was evaluated in a subgroup of participants by using between 2 and all acquired dynamics for the quantification. The corresponding cumulative maps} \deleted[id=(Rev2.14)]{$P_D^e$ maps} are displayed alongside each other together with plots from averaged signals from selected ROIs. \added[id=(Rev1-R1)]{\noindent In addition, a sliding window analysis was performed, whereby the perfusion map is the average over $l$ subsequent dynamics, where $l$ is the length of the sliding window.}

\subsection*{Additional Validation scans}
\noindent \added[id=(Rev2-R1)]{Additional experiments were performed to evaluate parameter choice, robustness and versatility of PERFOX} \deleted[id=(Rev2.14)]{Additional scans were acquired for}\added[id=(Rev2.14)]{on} \added[id=(Rev2-R1)]{a subset of subjects: (i) separate T$_2^*$ relaxometry, n=5; separate VSASL, n=5; and both separate T$_2^*$ and VSASL, n=1.} These separate scans were individually optimised for the respective purposes - narrow range of PLD for VSASL over a number of slices $N_s$ and ideal spacing of echo times for  T$_2^*$ measurements. Read-out parameters such as resolution, echo spacing and TR were chosen to be as close as possible between these acquisitions and the PERFOX acquisition to enhance consistency. The  T$_2^*$ acquisitions were performed in the same exam half immediately following the PERFOX scan; the VSASL scan was performed in the second exam half, after repositioning and new shimming. Reproducibility was studied in three volunteers where the PERFOX scan was repeated in the second half.\\

\noindent Finally, in eight of the 15 volunteers a PERFOX scan with higher resolution and axial slice orientation ({PERFOX-HIGH}) was performed\deleted[id=(Rev2.14)]{. The higher resolution was designed} to explore the ability of the proposed PERFOX technique to visualize even finer details. The different orientation was chosen as discussed above, to allow visualisation of the main functional \added[id=(Rev2.4)]{axis} of the placenta over a few select slices without the confounding effects of slice-dependent PLD and BGS variation. The TEs were chosen as [28,83,137] ms.\\

\section*{Results}
\subsection*{\deleted[id=(Rev2.10)]{Parameter optimisation}}
\begin{figure}[ht]
  \centering
  \scriptsize
  \includegraphics[width=0.9\textwidth]{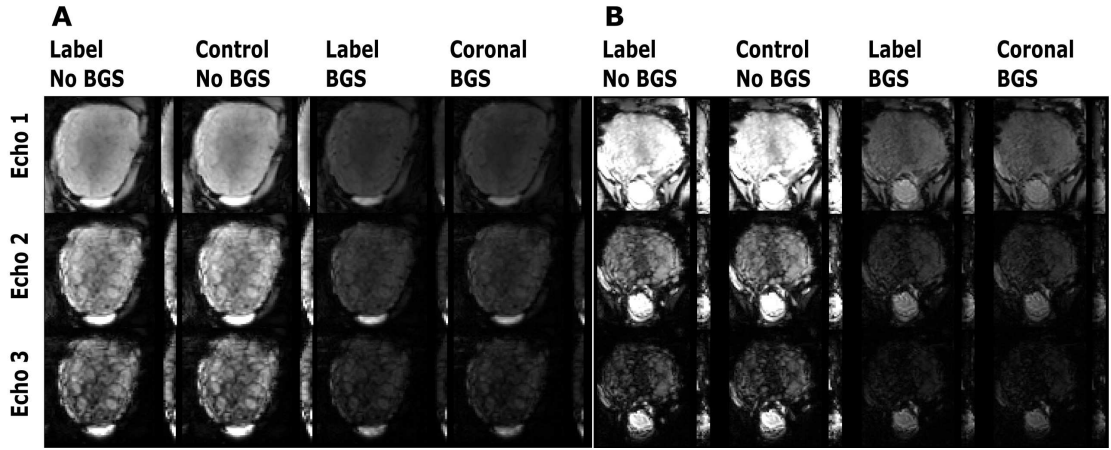}
  \caption{Depiction of \added[id=(Rev2-R1)]{two example PERFOX datasets from an (A) anterior and (B) posterior placenta. } All volumes are shown in a mid-parenchymal native coronal plane. Rows 1-3 show TE1-TE3 images.}
  \label{fig:dataraw} 
\end{figure}

\noindent The joint PERFOX scan was successfully implemented and acquired on all participants: \added[id=(Rev2-R1)]{\noindent unprocessed images from} \deleted[id=(Rev2.14)]{two PERFOX scans -} one anterior \added[id=(Rev2.7)]{placenta} and one posterior \added[id=(Rev2.7)]{placenta} are shown in Fig. \ref{fig:dataraw}. 

\deleted[id=(Rev2.11)]{The optimization of the joint scan required a reduction in the number of slices and number of TE, whilst making sure that  T$_2^*$ fitting was reliable and the PLD was not excessively long in the slices read-out last. The pragmatic choice of 8 slices resulted in a maximal PLD similar to the PLD in the 13th slice of a ('separate') VSASL acquisition with a single echo (i.e. no acquisition of simultaneous  T$_2^*$ relaxometry data). See Supporting Information Figure S1 for graphical illustration of the PLD of each slice and echo of the separate, PERFOX, and PERFOX-HIGH protocols. The constraints of the PERFOX acquisition, such as the coupling between inter-echo spacing and length of the EPI train guided the final choice of TEs of [{20},56,93] ms for the standard PERFOX acquisition, and [{28},83,137] ms for the high-resolution PERFOX-HIGH acquisition.}

\noindent \deleted[id=(Rev2.11)]{The BGS settings were based on preliminary T1 measurements of placenta tissue that showed T1 was between between 900 and 1200ms. Timings were optimised in order to keep placental tissue at approximately 10 percent of its equilibrium value at the excitation of the first slice readout \cite{Vidorreta2014}. This resulted in the two inversion pulses being placed right after the tagging module and 1130ms later. Supporting Information Figure S2 illustrates the effect of the choice of slice orientation on the acquired images, in terms of their differing dependence from PLD.}.

\subsection*{Motion correction}
\noindent \added[id=(Rev2.14)]{Visual analysis of a subset of the initial datasets confirmed that the registration of consecutive echoes was not beneficial: the extensive  T$_2^*$ differences between lobules and septa (tissue sections separating the lobules), result in different anatomical landmarks in the 3rd echo and frequent registration failures affecting also the alignment of the data from the shorter TEs. Better results were consistently achieved with the approach discussed above based on only estimating motion parameters from the volumes acquired at the first TE.}\\
\noindent \deleted[id=(Rev2.14)]{The two-step registration approach described in the Methods section, first registering all volumes acquired at the first echo time and then employing these same transformation for the subsequent echo times per volume followed the assumption that volumes from subsequent TEs share the same positions to the first TE volumes due to their temporal closeness (\textless 200 ms) and thus do not require further registration. This was verified in a subset of the initial datasets. This analysis confirmed that, as the 3rd TE provides less visible anatomical landmarks due to the extensive  T$_2^*$ differences between lobules and septa (tissue sections separating the lobules),  registrations involving the 3rd TE results in frequent registration failures affecting also the alignment of the data from the shorter TEs. Better results were consistently achieved with the approach discussed above based on only estimating motion parameters from first TE volumes.}\\

\begin{figure}[ht]
  \centering
  \scriptsize
  \includegraphics[width=0.9\textwidth]{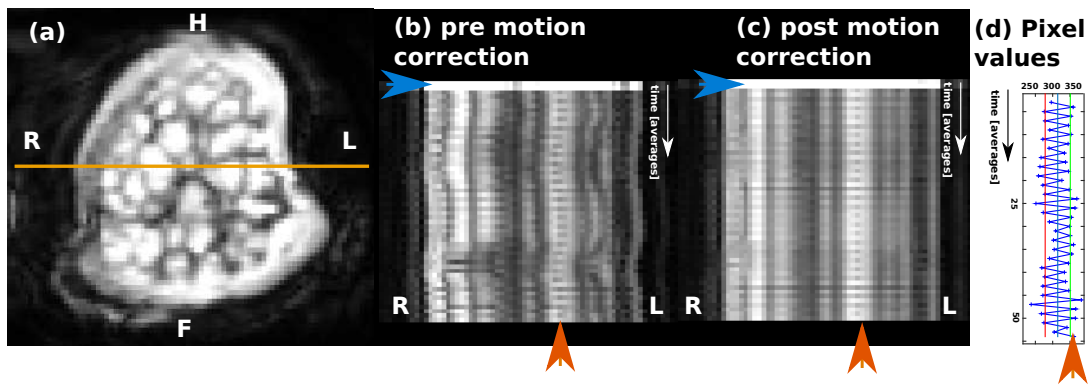}
  \caption{Illustration of the post-processing motion correction results. In (a) one mid-parenchymal slice in coronal orientation is shown to visualize the line in right-left orientation which is shown (b) before and (c) after motion correction over all volumes (top to bottom). The blue arrows indicate the first non background suppressed volumes, the orange arrow highlights an area of high perfusion, where the interleaved contrast is seen clearly before and after motion correction. (d) Finally, the post motion correction signal at the voxel depicted by the orange arrow in (b) and (c) is plotted together with the mean of the control volumes (green) and the mean of the labeled volumes (red).}
  \label{fig:moco}
\end{figure}

\noindent An example of the usefulness and efficacy of the motion correction is illustrated in Fig. \ref{fig:moco}, depicting one L-R line through the placenta for all dynamics (from top to bottom) pre- and post-motion correction (b-c). A better alignment is observed after motion correction (c), depicting both the alignment of similar structures but also, shown by orange arrows, the consistent signal changes from label to control on one select area of high perfusion. The depicted signal across the ROI (is shown in (d) corresponding to a signal mean of $280 \pm 17.08$ for the labeled and $350 \pm 12.36$ for the control volumes (compared to $305 \pm 31.02$ and $360 \pm 30.45$ pre-motion correction) - thus allowing for clear determination of the control-label signal difference.\\

\noindent The influence of the number of dynamics (control-label images) is illustrated in Fig. \ref{fig:dyn}, where all dynamics before motion correction are shown in the top row, the cumulative perfusion results \deleted[id=(Rev2.14)]{($P_D^e$) }and the sliding window perfusion results \deleted[id=(Rev2.14)]{($P_{Ds}^e$)} in the bottom row together with time curves for two voxels situated in high perfusion areas. \noindent Displaying this sliding window average across time gives an impression of the temporal variation of perfusion. \deleted[id=(Rev2.14)]{The results from three PERFOX scans, alongside the respective separately acquired VSASL and  T$_2^*$ are shown in Fig. \ref{fig:compare} displaying good qualitative and spatial agreement regarding the location of the high  T$_2^*$ and high perfusion areas.}\\

\subsection*{Spatial patterns}
\begin{figure}[ht]
  \centering
  \scriptsize
  \includegraphics[width=0.99\textwidth]{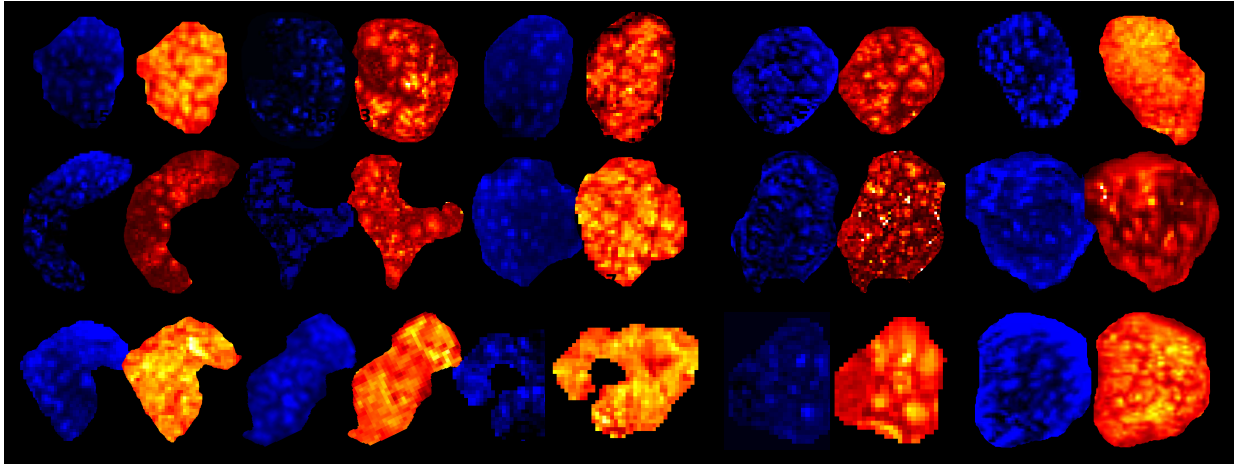}
  \caption{Overview over all 15 PERFOX datasets.  T$_2^*$ and Perfusion maps are given for the central slice. The colormaps are individually scaled from light blue-dark blue (high-low perfusion) and light yellow to dark red (high-low  T$_2^*$).}
  \label{fig:overview}
\end{figure}

\begin{figure}[ht]
  \centering
  \scriptsize
  \includegraphics[width=0.98\textwidth]{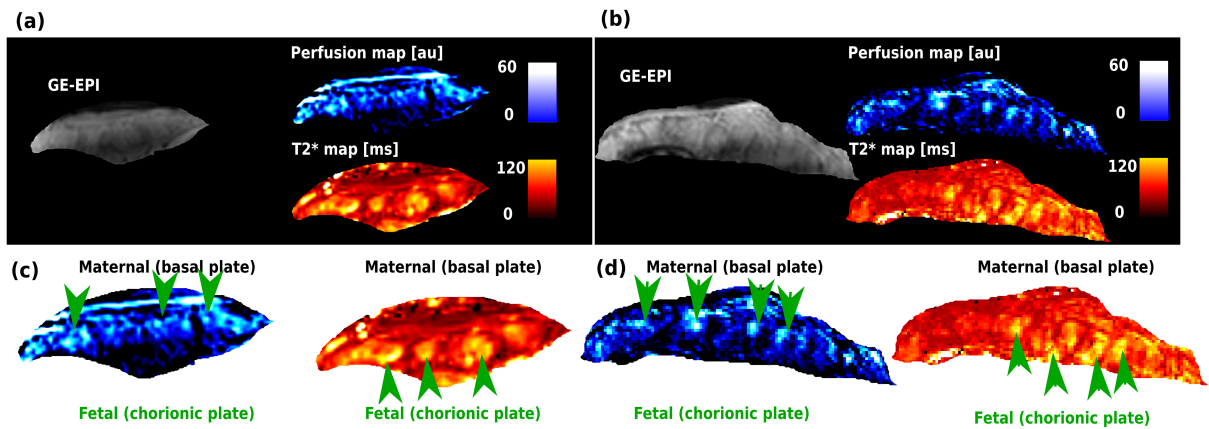}
 \caption{Perfusion and  T$_2^*$ maps obtained from the combined joint PERFOX-HIGH acquisition, in the axial plane, are shown on exemplary subjects. (a) GA 25+3 and (b) GA 38+1 weeks. The maps are shown separately overlayed on a Gradient-echo EPI image and then combined with the perfusion map overlayed on the  T$_2^*$ map. in (c-d) a zoom into the placental region is shown, the arrows indicate some of the areas of high  T$_2^*$ and high perfusion within the placental parenchyma.}
  \label{fig:combined_maps}
\end{figure}

\begin{figure}[ht]
  \centering
  \scriptsize
  \includegraphics[width=0.99\textwidth]{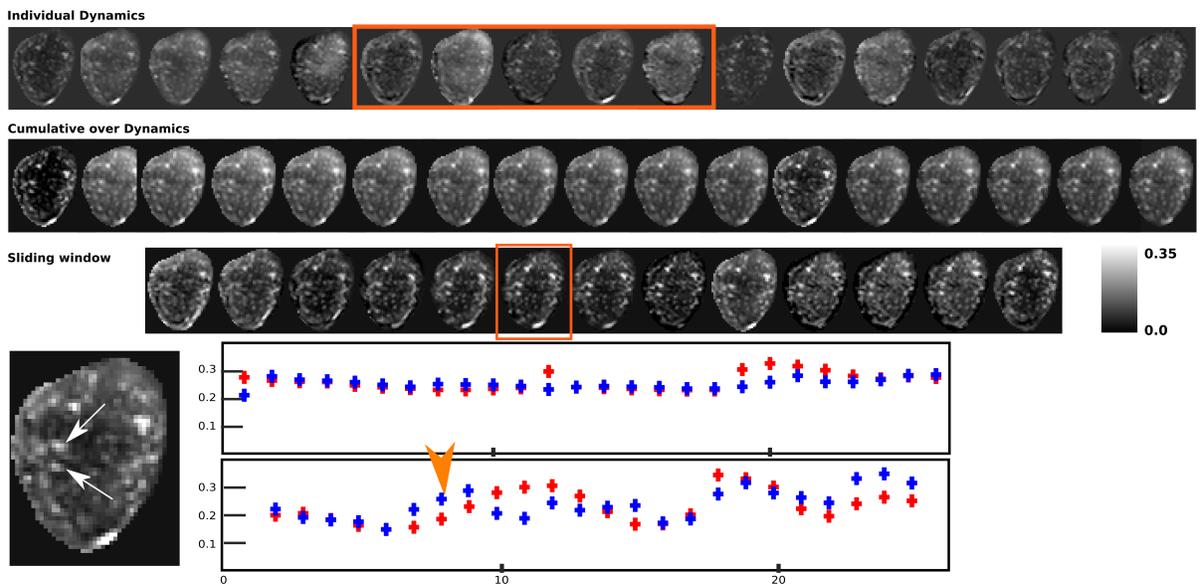}
  \caption{Results over dynamics. All dynamics are shown after subtraction in the top row, the results from cumulative analysis in the middle and the sliding window results with a window length of $l=5$ in the bottom row. Finally,  quantitative results from two voxels (red and blue) in high perfusion areas are depicted as a time curve for both analysis techniques (top: dynamics; bottom: sliding average).} 
  \label{fig:dyn}
\end{figure}

\begin{figure}[ht]
  \centering
  \scriptsize
  \includegraphics[width=0.9\textwidth]{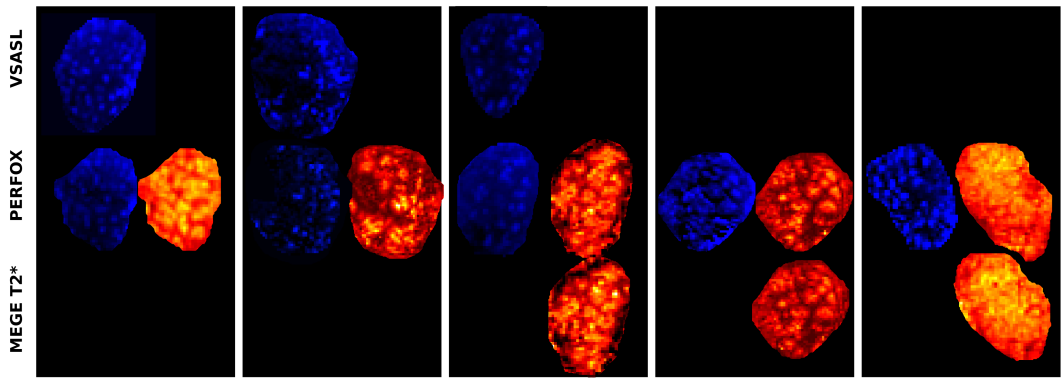}
  \caption{Results from PERFOX scans with matched separate VSASL (top row) and  T$_2^*$ (bottom row) scans. The colormaps are individually scaled from light blue-dark blue (high-low perfusion) and light yellow to dark red (high-low  T$_2^*$).} 
  \label{fig:compare}
\end{figure}

\noindent In the following,  T$_2^*$ results are consistently illustrated with red-yellow (low-high) colour scale and  perfusion results with \added[id=(Rev1-R1)]{dark blue - light blue (low-high) scale.} Resulting perfusion and  T$_2^*$ maps from one slice are given in Fig. \ref{fig:overview} for all participants and in Supporting Information Figure S3 for five slices of one participant. The images illustrate \added[id=(Rev2.12)]{ localized regions of high  T$_2^*$ and regions of high perfusion in the coronal planes. While their pattern is similar, the centers of these areas are not spatially co-localized within each slice}. The  T$_2^*$ maps show - in line with previous results \cite{Sorensen2013,Hutter2018} - multiple circular regions of variable size of long  T$_2^*$ with a clear peak in the middle and decay towards the outer regions. Perfusion weighted images and  T$_2^*$ maps are illustrated for the axial high resolution joint PERFOX-HIGH acquisition in Fig. \ref{fig:combined_maps} for two participants. The perfusion maps illustrate that the areas of highest perfusion appear close to the basal plate. These high perfusion regions then spread out branch-like from the maternal basal plate towards the chorionic plate. \deleted[id=(Rev2.14)]{In this axial view, it can be appreciated that the \added[id=(Rev1.1)]{areas of high-perfusion} are closer to the maternal basal plate and the} \added[id=(Rev2.14)]{The areas of high  T$_2^*$} are closer to the fetal chorionic plate (see \ref{fig:combined_maps} c-d). While only partial coverage of the placenta could be achieved in the transverse scans due to the longest axis of the placenta lying perpendicular to the visualised plane, these scans in this orientation best illustrate the non-co-localization of \added[id=(Rev1.1)]{areas of high  T$_2^*$ and perfusion} along the maternal-fetal axis. \added[id=(Rev2.14)]{The results from three PERFOX scans, alongside the respective separately acquired VSASL and  T$_2^*$ are shown in Fig. \ref{fig:compare}. They display good qualitative and spatial agreement regarding the location of the high  T$_2^*$ areas for PERFOX and the individual T$_2^*$ scan. Similarly, the areas of high perfusion areas acquired with PERFOX (middle row) and VSALS (top row) appear to correspond. However, their spatial alignment is less clear.}\\

\subsection*{Parameter evaluation and Quantitative group results}
\begin{figure}[ht]
  \centering
  \scriptsize
  \includegraphics[width=0.9\textwidth]{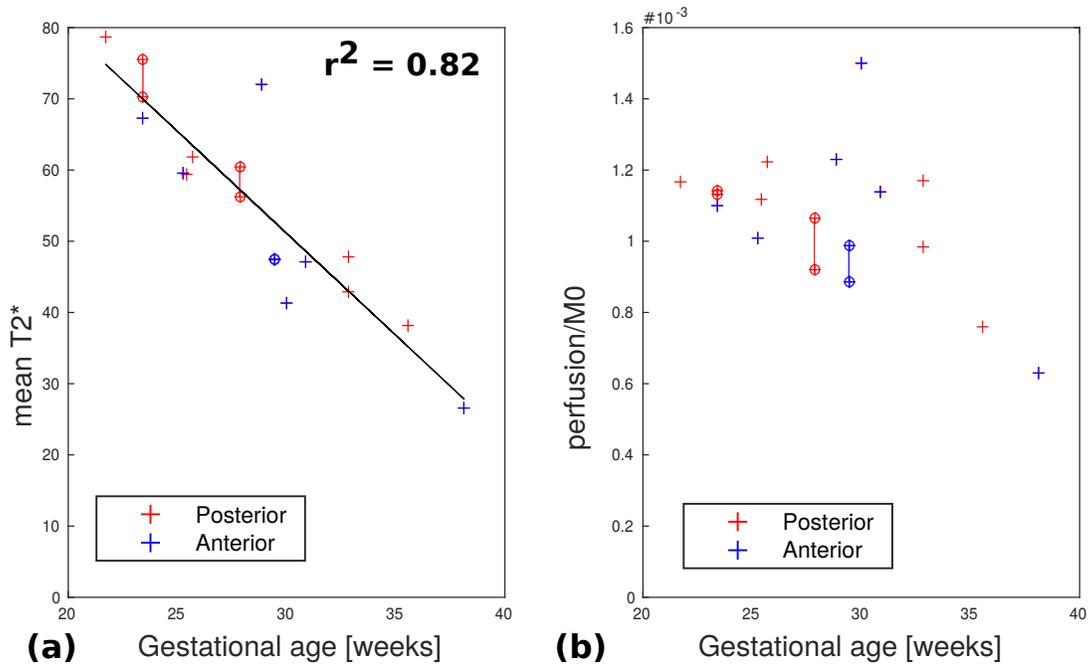}
  \caption{Quantitative results obtained for whole-placenta ROIs from all PERFOX scans. The mean  T$_2^*$ (a) and the mean perfusion/M0 (b) are depicted. The results from posterior placentas are shown in red, the results from anterior placentas are shown in blue. The repeated scans acquired for three subjects are highlighted by circles and connected by a line The decrease in perfusion with gestational age was not significant.}
  \label{fig:quantitative}
\end{figure}

\noindent Whole placental ROI analysis was performed on all \added[id=(Rev2.14)]{PERFOX scans.} \deleted[id=(Rev2.14)]{subjects who were acquired with the integrated PERFOX acquisition and the obtained quantitative results are depicted in Fig. \ref{fig:quantitative}.} The mean  T$_2^*$ over the whole organ is plotted against gestational age for all 18 scans in Fig. \ref{fig:quantitative}a, the perfusion/M0 results against gestational age are depicted in Fig. \ref{fig:quantitative}b. The points are colored by placental location, posterior in red and anterior in blue.
\added[id=(Rev1.4)]{\noindent T$_2^*$ and GA are significantly correlated (F=42.43, p\textless 0.05). There is no significant effect of placental location on  T$_2^*$ after controlling for GA (F=0.11, p=0.7426). There is no significant correlation between perfusion/M0 and GA (F=2.18, p=0.1484). There is also no significant effect of placental location on perfusion/M0 after controlling for GA (F=0.07, p=0.7992).} The results from the three participants with a repeated scan after re-positioning and new shimming are highlighted by circles and connected by lines. The mean coefficients of variation for these three repeated datasets are $4.6 \pm 1.5$ \% ( T$_2^*$) and $9.8 \pm 6.3$ \% (perfusion).\\ 

\noindent The multi-dimensional data obtained with PERFOX  allows several further directions of analysis: beside the conventional non-perfusion weighted  T$_2^*$ maps and the perfusion maps obtained with short TE, perfusion weighted  T$_2^*$ maps and perfusion maps at increasing levels of  T$_2^*$ weighting are shown in Supporting Information Figure S4. The  T$_2^*$ map from the control data shows longer  T$_2^*$ values compared to the tagged  T$_2^*$ map - illustrated as well in the difference map. Analysis of all datasets reveals \added[id=(Rev2.14)]{that} \deleted[id=(Rev2.14)]{similar behavior.} \added[id=(Rev2.14)]{the ratio between tagged and control  T$_2^*$, decreases over gestational age (p=0.0057). Similar analysis of the influence of the BGS on the T$_2^*$ maps did not show statistical significant differences between T$_2^*$ maps calculated from non-BGS and BGS volumes (p=0.88).}\\

\noindent The joint acquisition also allows to assess the effect of  T$_2^*$ weighting on the perfusion results by calculating the perfusion maps at different  T$_2^*$ weightings. The confounding effect of the  T$_2^*$ contribution to the perfusion map can be completely eliminated by calculating a perfusion map from the proton density maps extracted from  T$_2^*$ fitting. Example results comparing \added[id=(Rev2.14)]{the} perfusion map from the 1st TE and \deleted[id=(Rev2.14)]{from} the proton density maps are shown in Supporting Information Figure S5. They \deleted[id=(Rev2.14)]{. They look qualitatively very similar} are in good agreement for the central slices, but reveal differences which are localized mainly in the regions between the functional lobules.

\section*{Discussion}
\noindent This study presents a combined sequence for placental perfusion and oxygenation (PERFOX) measurement \deleted[id=(Rev2.14)]{in a combined sequence integrated PERFOX acquisition is presented} with required essential post-processing, mainly motion correction and quantification on 15 pregnant women. PERFOX is the first example of application of this dual-contrast acquisition to the placenta. A key advantage is the higher consistency within the multi-modal dataset, compared to a separate/sequential acquisition as well as the ability to dynamically resolve both essential quantities jointly. ASL measurements have, however, been previously combined with T2 \cite{Thomas2001,Wells2013} and diffusion measurements \cite{Silva1997,Wells2017}. The combination with  T$_2^*$ was originally exploited combining data from separate acquisitions at multiple echo times \cite{StLawrence2005} and then extended to dual-echo acquisitions, mainly for simultaneous measurements of cerebral blood flow and BOLD in fMRI \cite{Restom2007,Ghariq2014}.\\

\noindent The simultaneous acquisition allows to observe a clear spatial pattern in all our participants, with areas of high-perfusion centered close to the maternal basal plate and areas of high  T$_2^*$ closer to the fetal chorionic plate. This shift can be best observed in transverse scanning plane due to the curved geometry of the placenta with regard to the main imaging planes. \deleted[id=(Rev2.14)]{This can explain previous findings illustrating missing spatial co-localization of perfusion and  T$_2^*$ peaks \cite{Zhu2018}.} Furthermore, the low  T$_2^*$ on the in-flowing highly oxygenated maternal blood (identified by high perfusion signal) compared to the mid-parenchymal high- T$_2^*$ peaks might indicate that these observed high  T$_2^*$ regions are linked not only to oxygenation state but to either blood flow velocity, exchange or properties of fetal hemoglobin. The exact physiological pathway resulting in this observed behaviour remains however unclear, but dynamic multi-contrast techniques such as the acquisition presented here might help to shed light on these processes.\\
\added[id=(Rev2.13)]{Our choice of coronal acquisition plane as the main plane for the ongoing study, despite the above mentioned advantage of transverse scans to visualize the 'pathway' better, is driven by efficiency and continuity. As most placentas are either anterior or posterior with their main axis parallel to the foot-head direction, it can be covered by a lower number of slices in this direction compared to transverse. We appreciate however, that this is a choice and might not be the best solution for all purposes.}

\noindent  The analysis \added[id=(Rev1-R1)]{provided here, showed that the number of required dynamics for stable perfusion signal in the placenta is in the range of 5-8 for the chosen acquisition parameters.} \deleted[id=(Rev1.3a)]{This allows either for reduction in the number of dynamics, to shorten the required acquisition time further or to focus on dynamic assessment. The data presented towards the latter illustrated results from subsequent dynamics in a sliding window type analysis.} \added[id=(Rev1.3a)]{This allows either to reduce the number of dynamics and thus limit the required acquisition time, or a dynamic assessment using a sliding window analysis as illustrated in Figure \ref{fig:dyn}.}

\deleted[id=(Rev1.3b)]{This is a first step to make ideal use of the dynamic PERFOX acquisition and has clear benefits with regard to the expected motion range and dynamic nature of effects such as contractions in the chosen application.} 
\deleted[id=(Rev1.3c)]{The high contrast and perfusion values observed in the placenta are essential for this.} \added[id=(Rev1.3c)]{The high contrast to noise ratio resulting from high perfusion values observed in the placenta are essential for this.} These observations will not directly translate to less blood-rich organs such as the brain, \deleted[id=(Rev2.14)]{ Obviously} \added[id=(Rev2.14)]{where a higher number of dynamics is required for robust perfusion visualization, resulting in coarser temporal resolution.}

\noindent The observed strong linear decrease over gestation in  T$_2^*$ and weak negative correlation between perfusion and gestation are in line with those observed separately in previous studies \cite{Zun2017,Sorensen2013,Sinding2017}.  Zun et al. \cite{Zun2017} reported higher perfusion for posterior placentas. Our data show no such significant difference based on placental location. There is little evidence of differences in placental function between anterior and posterior placental locations in literature, and our results would support this. The low number of participants in the current placental ASL studies however call for caution regarding both differences with location and trends over gestation.

\noindent One prior study showed differences in perfusion between lateral and supine positioning \cite{Zun2017}. In contrast, at our institution all pregnant participants are scanned in supine position \added[id=(Rev2-R1)]{while under constant monitoring and splitting the scanning time}. \deleted[id=(Rev2-R1)]{Whilst some investigators have warned against the risk of vena cava compression syndrome in this position, a recent study found that the spinal venous plexus and ascending lumbar veins act as a venous return system and allow the maintenance of vascular homeostasis in pregnant women lying supine; this suggests that scanning pregnant women in the supine position is safe \cite{Hughes2016}. Using such positioning consistently allows to increase inter-subject comparability and consistency of the quantitative data.}

\noindent Recent placental ASL studies used 3D readouts \cite{Zun2017,Shao2018}\deleted[id=(Rev2.14)]{, In contrast 2D EPI acquisition was chosen here} \added[id=(Rev2.14)]{in contrast to the 2D EPI acquisition chosen for this study}. We selected 2D EPI due to its ability to freeze motion within each slice, the flexibility to optimize the echo spacing in order to reduce acoustic noise \deleted[id=(Rev2.14)]{to the scanner specific frequency transfer functions. This choice furthermore allows} and to acquire the data required for  T$_2^*$ relaxometry with multiple echoes in quick succession. \\

\noindent Compared with more conventional \added[id=(Rev1-R1)]{separate} acquisitions which allow individual optimisation of sequence parameters for each modality - e.g. 4 echo times spanning a wide range for  T$_2^*$ mapping and narrow range of PLDs for all slices - the joint acquisition inevitably forces compromises on these constraints. Therefore, i.e. only 3 echo times for  T$_2^*$ mapping with lower maximal TE of 93ms (instead of 148ms) and a \added[id=(Rev2.14)]{reduced} coverage of 8 slices (instead of 13) were chosen to keep the acquisition time for all slices at all three echo times as compact as possible. Nevertheless, the obtained functional maps illustrate usable data of good quality. \added[id=(Rev2.5)]{The Rician noise distribution in conventional MRI images, approaching Gaussian distribution only for high signal-to-noise ratio (SNR), is important to consider for the  T$_2^*$ fits. Low SNR is associated with longer echo times, which result e.g. from longer EPI trains to achieve higher resolution multi-echo  T$_2^*$ scans or if more then 4 echoes are acquired for multi-exponential multi-compartment fits. However, in our case, the requirement to reduce PLD resulted in only three echo times acquired at comparably low resolution of 3mm. The short TEs ensure robust mono-exponential fitting, but do, however, not support higher order fitting.}

\deleted[id=(Rev2.6)]{Velocity-selective ASL was chosen above other types of ASL due to its spatially non-selective, and thus easily applicable tagging module. This facilitates accurate and robust tagging in the placenta where the supplying vasculature and region of interest are highly variable.} VSASL often employs a second velocity selective module immediately before the acquisition (this sequence is referred to as dual VSASL or DVASL in \cite{Schmid2015}): it saturates blood flowing above the cutoff velocity and thus acts as a filter to make sure signal contributions from blood accelerating during the PLD are eliminated (i.e. in the brain this would be venous blood); without this second module 
VSASL images are very difficult to quantify as they have contributions from both arterial and venous flow \cite{Schmid2015}.
\added[id=(Rev2.14)]{However, for this study VSASL with a single VS module was chosen} for a number of reasons. Firstly, the complex placental vasculature with slow-flowing oxygen-rich blood between the villi and oxygen-poorer and higher-velocity venous backflow \deleted[id=(Rev2.14)]{might not be ideally suited for this approach; indeed, blood on the fetal side, accelerating from the villi towards the fetus might be of interest for placental application} \added[id=(Rev2.14)]{does not allow a clear velocity-based separation into arterial and venous blood.} Secondly, each of the encoding modules leads to substantial T2-dependent signal decay.  With our 50ms-long module and assuming T2 of blood to be roughly 170ms, we have a 35.5\% signal loss/module. Less oxygenated blood with lower T2 will experience an even higher signal reduction. Finally, the stricter specific absorption rate (SAR) limitation for fetal studies puts a time penalty on the DVSASL technique where 2 additional 90-degree pulses plus an adiabatic refocusing pulse are required. In our case adding the 2nd module raised the minimally achievable repetition time from 3.5s to 6.4s. This would thus either increase the acquisition time or decrease the number of acquired label-control pairs. Evaluation of dual VSASL vs the protocol used in this study is underway.\\

\noindent This study does not describe variations of the velocity encoding direction; head to feet encoding was chosen for all scans \added[id=(Rev2.6)]{irrespective of scan orientation}, similar to \cite{Zun2017}. The results reported appear to be consistent with labelling of maternal blood. Experiments are ongoing to further quantify the effect of velocity encoding in different directions. One of the previous placental VSASL studies \cite{Zun2018} reported a within-subject coefficient of variation of only 3.5\% on whole-placenta perfusion values. Whilst the coefficient of variation reported here is higher at 9.8\%, it is important to note that we assessed reproducibility \added[id=(Rev2.14)]{between two sessions,} \deleted[id=(Rev2.14)]{, including repositioning,} providing a much more appropriate estimate of data reliability for a clinically useful scanning scenario than the back-to-back scanning reported in\cite{Zun2018}.\\

\added[id=(Rev1-R1)]{\noindent Six separate conventional VSASL and T$_2^*$ scans were acquired in a subgroup of subjects. However, whilst the  T$_2^*$ scans were acquired immediately before or after the PERFOX scans, within the same session, the separate perfusion scans were acquired in different scanning sessions due to the restrictions on continuous scanning time for pregnant women. This difference is reflected both qualitatively and quantitatively: while both show visually good agreement regarding the location and size of the areas of high perfusion and  T$_2^*$, it was significantly harder to find the same geometrical location for the perfusion maps acquired in different scanning sessions due to changes in maternal positioning, fetal lie, location of the fundus and planning of the region of interest for the acquisition. Future work will include validation with a static perfusion phantom.}\\

\noindent In this study, the joint-acquisition data was presented separately by processing along the echo times for relaxometry and in a separate step along dynamics for perfusion information. However, the data can be processed to show differences in the  T$_2^*$ maps calculated from control and labeled volumes; the possibility to calculate the perfusion map from the proton density map - allowing to correct for  T$_2^*$ effects - is also appealing. Both are additional benefits of the joint acquisition. Indeed, the data is ideally suited for a fully combined analysis approach, that will be explored in the future.\\

\noindent This study proposes \added[id=(Rev1-R1)]{a novel combined strategy} to obtain the co-localized functional descriptor \deleted[id=(Rev2.14)]{to allow visualisation of two important elements relating to placental function -} \added[id=(Rev2.14)]{visualizing} perfusion and oxygenation. The PERFOX joint acquisition will be used in the future to study a variety of functional mechanisms such as the causality of insufficiency \added[id=(Rev2.14)]{ in pregnancy complications like pre-eclampsia, growth restriction and congenital heart disease. Visualizing} an imbalance between perfusion and oxygen uptake\deleted[id=(Rev2.14)]{. This} might allow to explore compensatory mechanisms and to study variations over placental surface and thus possibly deviations in implantation. \deleted[id=(Rev2.14)]{Furthermore, it will help shed light on physiological changes in pregnancies complicated by pre-eclampsia, fetal growth restriction and fetal congenital heart disease, e.g. to assess the direction of changes - limited perfusion or limited oxygen uptake from the fetal circulation.} This might ultimately benefit the deeper understanding of placental physiology and disease aetiology. \deleted[id=(Rev2.14)]{in pregnancy complications.} Lastly, the proposed joint PERFOX sequence might furthermore find applications in other, highly perfused organ systems such as kidney or liver where an equal interest exists with regard to separating perfusion and oxygenation effects.

\newpage 
\section*{Acknowledgments}
We thank the midwives, obstetricians and radiographers who played a key role in obtaining the data sets. We would also like to thank all participating mothers. This work received funding from the NIH  (Human Placenta Project - grant 1U01HD087202-01), the Wellcome Trust (Sir Henry Wellcome Fellowship, 201374/Z/16/Z), and the EPSRC (grants N018702 and M020533). This work was also supported by the Wellcome/EPSRC Centre for Medical Engineering [WT 203148/Z/16/Z].

\bibliography{t2sasl}
\newpage

Figure 1: The placenta is depicted schematically. The spiral arteries ensuring supply from maternal side, the villous trees for oxygen uptake as well as a (schematic) depiction of the oxygen concentration from red (high oxygen content) to blue (low oxygen content) are illustrated together with the definition for a functional lobule used in the following. Furthermore, dotted lines point to the basal and chorionic plate. The right side shows some of the hypothesized changes in pre-eclampsia: elongated, less-capillarized trees and increased thickness.\\
  
Figure 2: The acquisition strategy (A) and post-processing (B) for PERFOX are depicted schematically. In (A) the VSASL labelling module, the background suppression module and the acquisition module - composed of three echoes - are depicted. NB. the horizontal time axis is not to scale.\\
  
Figure 3: Depiction of two example PERFOX datasets from an (A) anterior and (B) posterior placenta. All volumes are shown in a mid-parenchymal native coronal plane. Rows 1-3 show TE1-TE3 images.\\
  
Figure 4: Illustration of the post-processing motion correction results. In (a) one mid-parenchymal slice in coronal orientation is shown to visualize the line in right-left orientation which is shown (b) before and (c) after motion correction over all volumes (top to bottom). The blue arrows indicate the first non background suppressed volumes, the orange arrow highlights an area of high perfusion, where the interleaved contrast is seen clearly before and after motion correction. (d) Finally, the post motion correction signal at the voxel depicted by the orange arrow in (b) and (c) is plotted together with the mean of the control volumes (green) and the mean of the labeled volumes (red).\\
 
 Figure 5: Overview over all 15 PERFOX datasets.  T$_2^*$ and Perfusion maps are given for the central slice. The colormaps are individually scaled from light blue-dark blue (high-low perfusion) and light yellow to dark red (high-low  T$_2^*$).\\
  
 Figure 6: Perfusion and  T$_2^*$ maps obtained from the combined joint PERFOX-HIGH acquisition, in the axial plane, are shown on exemplary subjects. (a) GA 25+3 and (b) GA 38+1 weeks. The maps are shown separately overlayed on a Gradient-echo EPI image and then combined with the perfusion map overlayed on the  T$_2^*$ map. in (c-d) a zoom into the placental region is shown, the arrows indicate some of the areas of high  T$_2^*$ and high perfusion within the placental parenchyma.\\
 
 Figure 7: Results over dynamics. All dynamics are shown after subtraction in the top row, the results from cumulative analysis in the middle and the sliding window results with a window length of $l=5$ in the bottom row. Finally,  quantitative results from two voxels (red and blue) in high perfusion areas are depicted as a time curve for both analysis techniques (top: dynamics; bottom: sliding average).\\

Figure 8: Results from PERFOX scans with matched separate VSASL (top row) and  T$_2^*$ (bottom row) scans. The colormaps are individually scaled from light blue-dark blue (high-low perfusion) and light yellow to dark red (high-low  T$_2^*$).\\

Figure 9: Quantitative results obtained for whole-placenta ROIs from all PERFOX scans. The mean  T$_2^*$ (a) and the mean perfusion/M0 (b) are depicted. The results from posterior placentas are shown in red, the results from anterior placentas are shown in blue. The repeated scans acquired for three subjects are highlighted by circles and connected by a line The decrease in perfusion with gestational age was not significant.\\

Supporting Information Figure S1: Post labeling delays illustrated for all slices and echoes for both 'separate' VSASL with 13 slices (blue)) and the two PERFOX variants with 8 slices used in the paper (standard PERFOX in green and PERFOX-HIGH in red). The first echo is marked by a large star, second and third echoes by smaller stars.\\

Supporting Information Figure S2: Illustration of the influence of the different scan orientations on the acquired signal. (a) Schematic illustration of the two scan orientations employed in this paper - coronal and transverse. (b-c) Control images from the coronal (red background) and transverse (green background) acquisitions, each is also displayed reformatted in the non-native orientation. (b) displays the results without and (c) with background suppression. Finally, (d) displays a zoom into both acquisitions together with yellow arrows to illustrate the direction of increasing PLD.\\

Supporting Information Figure S3: Results from a coronal PERFOX scan at GA 29+1 weeks. Five consecutive slices are shown for the anatomical GE-EPI volume (first row), the perfusion maps (second row) and the  T$_2^*$ maps (third row).\\

Supporting Information Figure S4: (a)  T$_2^*$ maps calculated from the control volumes, labelled volumes  and difference in  T$_2^*$ between the two.(b) Perfusion maps at the 3 different echo times TE acquired in PERFOX. (c) Evaluation over all subjects of the mean  T$_2^*$ values from tagged and control volumes relative to the  mean  T$_2^*$ from control volumes.\\

Supporting Information Figure S5: Perfusion maps obtained from the proton density maps and from the data from the 1st echo time together with difference image.\\
\newpage 

\newpage

\end{document}